\documentclass[aps,prb,reprint, superscriptaddress,secnumarabic,amssymb,nobibnotes]{revtex4-1}
\usepackage{graphicx}
\usepackage{amsmath}
\usepackage{subfigure}
\usepackage{verbatim}
\usepackage[utf8]{inputenc}
\usepackage{dsfont}
\usepackage[colorinlistoftodos]{todonotes}
\usepackage[version=3]{mhchem}
\usepackage{chemfig}
\usepackage{cancel}
\usepackage{hyperref}
\usepackage{xcolor}
\usepackage{upgreek}
\usepackage{bm}
\usepackage{amssymb}

\makeatletter
\newsavebox{\@brx}
\newcommand{\llangle}[1][]{\savebox{\@brx}{\(\m@th{#1\langle}\)}%
\mathopen{\copy\@brx\kern-0.5\wd\@brx\usebox{\@brx}}}
\newcommand{\rrangle}[1][]{\savebox{\@brx}{\(\m@th{#1\rangle}\)}%
\mathclose{\copy\@brx\kern-0.5\wd\@brx\usebox{\@brx}}}
\makeatother

\begin{document}

\title{Nonreciprocal ballistic transport in asymmetric bands}

\author{M. H. Zou}
\affiliation{National Laboratory of Solid State Microstructures, School of Physics,
and Collaborative Innovation Center of Advanced Microstructures, Nanjing University, Nanjing 210093, China}

\author{H. Geng}
\email{genghao@nju.edu.cn}
\affiliation{National Laboratory of Solid State Microstructures, School of Physics,
and Collaborative Innovation Center of Advanced Microstructures, Nanjing University, Nanjing 210093, China}

\author{R. Ma}
\affiliation{
Nanjing University of Information Science and Technology, Nanjing 210044, China}

\author{Wei Chen}
\affiliation{National Laboratory of Solid State Microstructures, School of Physics,
and Collaborative Innovation Center of Advanced Microstructures, Nanjing University, Nanjing 210093, China}

\author{L. Sheng}
\email{shengli@nju.edu.cn}
\affiliation{National Laboratory of Solid State Microstructures, School of Physics,
and Collaborative Innovation Center of Advanced Microstructures, Nanjing University, Nanjing 210093, China}

\author{D. Y. Xing}
\affiliation{National Laboratory of Solid State Microstructures, School of Physics,
and Collaborative Innovation Center of Advanced Microstructures, Nanjing University, Nanjing 210093, China}


\date{\today }

\begin{abstract}
Nonreciprocal transport in uniform systems has attracted great research interest recently
and the existing theories mainly focus on the diffusive regime.
In this study, we uncover a novel scenario for nonreciprocal charge transport in the ballistic regime enabled by
asymmetric band structures of the system. The asymmetry of the bands induces unequal 
Coulomb potentials within the system as the bias voltage imposed by the electrodes inverts its sign.
As a result, the bands undergo different energy shifts as the current flows in opposite directions,
giving rise to the nonreciprocity. Utilizing the gauge-invariant nonlinear transport theory, 
we show that the nonreciprocal transport predominantly originates from the second-order conductance,
which violates the Onsager reciprocal relation but fulfills a generalized reciprocal relation similar to that of unidirectional magnetoresistance.
The ballistic nonreciprocal transport phenomena differ from the diffusive ones by considering the internal asymmetric Coulomb potential, a factor not accounted for in diffusive cases but undeniably crucial in ballistic scenarios.
Our work opens a avenue for implementing nonreciprocal transport in the ballistic regime and
provides an alternative perspective for further experimental explorations for nonreciprocal transport.
\end{abstract}

\maketitle

\section{Introduction}
Reciprocity in charge transport reflects a symmetrical relationship
between current and voltage, where the magnitude of current stays constant when
the voltage has an opposite sign but the same value~\cite{Tokura2018, Ideue2021}.
Violations of reciprocity underpin the functionality of key electronic devices like diodes and photodetectors~\cite{Fruchart2021,Akamatsu2021}.
Although nonreciprocal transport is commonly encountered at systems involving interfaces,
such as the celebrated p-n junction,
its implementation in uniform bulk materials has occurred more recently, driven by entirely different scenarios.
A novel type of nonreciprocal transport, termed electric magnetochiral anisotropy (EMCA), which
exhibits unidirectional magnetoresistance (UMR), was first proposed in Ref.~\onlinecite{Rikken2001}.
Inspired by this discovery, extensive exploration has been conducted to identify material candidates
that exhibit nonreciprocal transport. Significant findings encompass chiral
nanosystems~\cite{Rikken2001, Krstic2002}, polar semiconductors~\cite{Ideue2017,Itahashi2020, Li2021},
bilayer heterojunctions~\cite{Avci2015, Yasuda2019,Choe2019, Shim2022, Ye2022}, and topological systems~\cite{Yasuda2020,Zhang2022}.

On the theoretical side, a variety of mechanisms have been
put forward to explain nonreciprocal transport in systems with translational symmetry. 
Typical scenarios include asymmetric band structures~\cite{Tokura2018,Ideue2017}, asymmetric inelastic scattering by spin clusters~\cite{Ishizuka2020}
and magnons~\cite{Yasuda2016}, quantum metric~\cite{Kaplan2022, Wang2023}, and non-Hermitian skin effect~\cite{Geng2023,Shao2023}.
In these studies, various theoretical approaches have been employed, such as semiclassical transport equations~\cite{Zhang2016, Ideue2017,Sterk2019,Kaplan2022} and quantum kinetic theory~\cite{Freimuth2021}, both of which consider the nonlinear effects of an external electric field.
Here, we mainly focus on the nonreciprocal transport in the systems with asymmetric bands.
The existing mechanism for this case mainly concentrates on the diffusive limit\cite{Ideue2017}, and a direct extension to the ballistic limit is not feasible
under the same theoretical framework. 
Therefore, it remains an open question
that whether nonreciprocal transport can be achieved in the ballistic regime.

\begin{figure}[t]
    \centering
    \includegraphics[width=1\columnwidth]{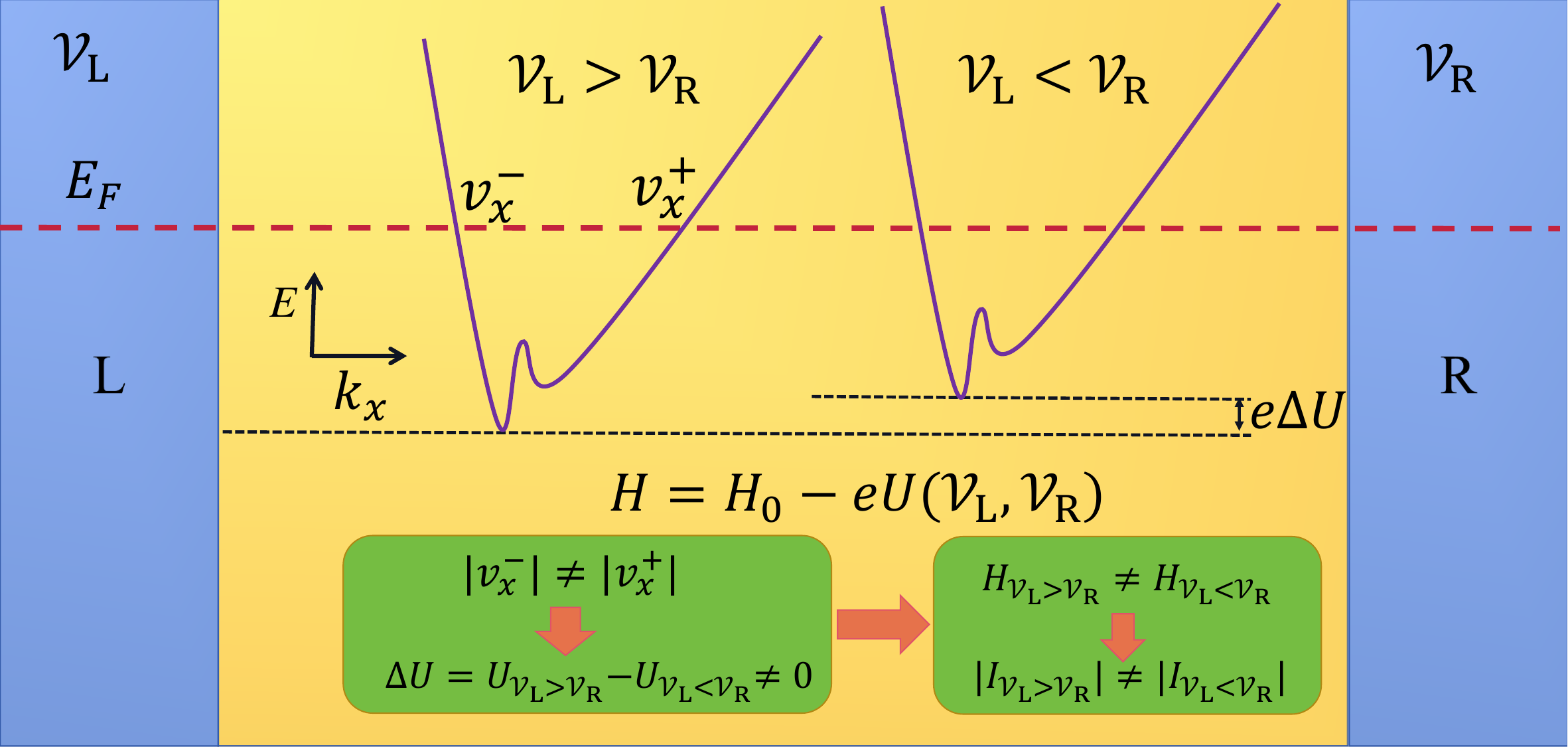}
    \caption{The two-terminal setup for the nonlinear nonreciprocal transport system with asymmetric band structures. The terminals, depicted in blue, are designated as ‘L’ (left) and ‘R’ (right) with biases \( \mathcal{V}_{\text{L}} \) and \( \mathcal{V}_{\text{R}} \). The region in yellow represents the scattering regime. The asymmetric band structure is illustrated by the purple lines. In equilibrium, the Fermi level and the Hamiltonian are denoted by \( E_F \) and \( H_0 \), respectively. The internal screened Coulomb potential, represented as \( U(\mathcal{V}_{\text{L}},\mathcal{V}_{\text{R}}) \), correlates with the biases at the terminals. The term \( v^{\pm}_x \) indicates the velocity of the right-moving (left-moving) mode around the Fermi energy. The current generated in the system, as a result of the potential difference, is expressed by \( I(\mathcal{V}_{\text{L}},\mathcal{V}_{\text{R}}) \). 
    $\Delta U$ the difference of Coulomb potential $U(\mathcal{V}_{\text{L}},\mathcal{V}_{\text{R}}) $ in the cases of $\mathcal{V}_{\text{L}}>\mathcal{V}_{\text{R}}$ and $\mathcal{V}_{\text{L}}<\mathcal{V}_{\text{R}}$, makes a shift of the bottom of energy bands.
    }\label{fig0}
\end{figure}

In this work, we give an affirmative answer to this question by showing that nonreciprocal ballistic transport
can be implemented in systems featuring asymmetric band structures with the setup depicted in Fig. \ref{fig0}. 
In this scenario,
the asymmetry of the bands induces unequal Coulomb potentials \( U(\Delta V) \) within the system as the bias difference $\Delta V = \mathcal{V}_{\text{L}}-\mathcal{V}_{\text{R}}$ between the terminals
undergoes a change in sign, because the current flowing in opposite directions
is carried by electrons with different densities. This leads to different energy shifts as the current flows in opposite directions.
Consequently, the states occupied by electrons in the bands are situated at different levels in the
original band structures without the effect by Coulomb potentials, resulting in nonreciprocal transport.
By utilizing the theoretical framework of gauge-invariant nonlinear quantum transport~\cite{Buettiker1993,Buettiker1995, Christen1996}
we demonstrate the nonreciprocal ballistic transport in a two-terminal setup by showing \(|I(\Delta V, B)| \neq |I(-\Delta V, B)|\),
where $B$ is the magnetic field utilized to generate band asymmetry.
We further show that although the Onsager reciprocal relation\cite{Onsager1931}
breaks down here, the relation \(|I(\Delta V, B)| = |I(-\Delta V, -B)|\) similar to that in EMCA~\cite{Rikken2001,Rikken2005} maintains.

The rest of the paper is organized as follows:
In Section II, we provide a concise review
of the theory employed in our study.
In Section III a general theory of nonreciprocal ballistic
transport is elucidated.
In Section IV, the physical results are specified in the 2D Rashba gas
subjected to a in-plane magnetic field.
Finally, in Section V, we
present our concluding remarks.

\section{Gauge Invariant Nonlinear Quantum Transport Theory}
Consider quantum coherent transport taking place
in mesoscopic system with connection to multiple terminals labeled by \( \{\alpha\} \).
The electric current \( I_\alpha \) in terminal \( \alpha \)
driven by the bias voltages \( \{\mathcal{V}_{\alpha}\} \)
is expressed as~\cite{Jauho1994,Anantram1995,Christen1996,Datta2005}:
\begin{equation}\label{Currenteq}
    \begin{split}
    I_{\alpha}
    &=-\frac{e}{h}\sum_{\beta}\int dE
    T_{\alpha \beta}\left(E, U\right)\left(f_{\alpha}-f_{\beta}\right),
    \end{split}
\end{equation}
where \( f_{\alpha}\equiv f\left(E-\mu_{\alpha}\right) \) is the Fermi-Dirac distribution function in terminal \( \alpha \), with \( \mu_{\alpha} = E_F - e\mathcal{V}_{\alpha} \), \( E_F \) the equilibrium Fermi energy, and \( \mathcal{V}_{\alpha} \) the bias voltage.
The transmission from terminal \( \beta \) to \( \alpha \) is given by \( T_{\alpha \beta} = \text{Tr}\left[\Gamma_{\alpha}\mathcal{G}^r\Gamma_{\beta}\mathcal{G}^{a}\right] \), where
\( \mathcal{G}^{r(a)} \) is the retarded (advanced) Green's function defined as
\begin{equation}\label{retgreeneq}
    \begin{split}
        \mathcal{G}^{r(a)}(E,U) &= \frac{1}{E-H+eU-\Sigma^{r(a)}},\\
        \Sigma^{r(a)}&=\sum_\alpha \Sigma_\alpha^{r(a)},
    \end{split}
\end{equation}
with \( \Sigma_{\alpha}^{r(a)} \) the retarded (advanced) self-energy introduced by terminal \( \alpha \)
that satisfy $\Sigma_\alpha^a=(\Sigma_\alpha^r)^\dag$.
The linewidth function is defined as \( \Gamma_{\alpha} = i\left(\Sigma_{\alpha}^{r} - \Sigma_{\alpha}^{a}\right) \).
For a two-terminal setup, the unitarity of the scattering matrix~\cite{Bruus2004} ensures \( T_{LR} = T_{RL} \).

In Eq.~\eqref{retgreeneq}, \( H \) is the Hamiltonian of the system in equilibrium,
i.e., all bias voltages vanish (\( \mathcal{V}_{\alpha} = 0 \)).
The additional term \( U(x) \) is the Coulomb potential arising from a finite bias,
which satisfies the Poisson equation~\cite{Wang1999}
\begin{equation}\label{poissoneq}
    \nabla^2 U(x) = 4 \pi i e \int\frac{dE}{2\pi}\left[\mathcal{G}^{<}(E, U)\right]_{xx},
\end{equation}
where \( x \) denotes the position. The potential $U(x)$ plays an essential
role for the nonreciprocal ballistic transport as the asymmetric band structures
are considered. The lesser Green's function \( \mathcal{G}^{<} \) is defined as
\( \mathcal{G}^{<} = \mathcal{G}^{r} \Sigma^{<} \mathcal{G}^{a} \), with
\begin{equation}\label{lesssigmaeq}
    \Sigma^{<} = \sum_{\alpha} i \Gamma_{\alpha} f_{\alpha}.
\end{equation}

In general, a self-consistent approach is required to solve Eqs.~\eqref{retgreeneq}-\eqref{lesssigmaeq}
in the nonlinear regime. Since the lesser Green's function exhibits a nonlinear relationship
with \( U \), Eq.~\eqref{poissoneq} is a nonlinear differential equation.
Nevertheless, the entire theoretical framework is gauge-invariant\cite{Wang1999},
which means that the current is invariant under a uniform potential shift applied throughout the system.

Here, we focus on the weakly nonlinear regime, where the Coulomb potential can be expanded as
\begin{equation}\label{Uexpandeq}
    U(x) = \sum_{\alpha} u_{\alpha}(x)
    \mathcal{V}_{\alpha} + \cdots,
\end{equation}
where the zeroth-order term (potential in equilibrium) has been absorbed into
the Hamiltonian $H$, and \( u_{\alpha}(x) \) denotes the characteristic potential~\cite{Christen1996,Wang1999}.
Gauge invariance of the theory requires~\cite{Christen1996}
\begin{equation}
    \sum_{\alpha} u_{\alpha} = 1.
\end{equation}

To the lowest order, we derive the equation for \( u_{\alpha}(x) \) from Eqs.~\eqref{poissoneq} and \eqref{Uexpandeq}~\cite{Buttiker1993,Christen1996} as
\begin{equation}\label{cpueq}
    - \nabla^2_x u_{\alpha}
    +4 \pi e^2 \frac{dn}{dE} u_{\alpha}
    = 4\pi e^2 \frac{dn_{\alpha}}{dE},
\end{equation}
where $n(x)$ is the local charge density,
and the injectivity or local partial density of states (LPDOS) of
terminal \( \alpha \)\cite{Buttiker1993,Christen1996} is given by
\begin{equation}
    \frac{dn_{\alpha}}{dE} (x)=\int\frac{dE}{2\pi}
    (-\partial_E f_0)
    \left[
        \mathcal{G}_0^r\Gamma_{\alpha} \mathcal{G}_0^a
    \right]_{xx},
\end{equation}
in which \( \mathcal{G}_0^{r,a} \) is the equilibrium Green's function with \( U(x) = 0 \).
An alternative expression for injectivity in terms of scattering wavefunctions and the velocity of incident modes is\cite{Wang1997,Kramer2012}
\begin{equation}\label{LPDOSPSiEq}
    \frac{dn_{\alpha}}{dE} (x) = \int\frac{dE}{2\pi}
    (-\partial_E f_0)\sum_n
    \frac{|\Psi_{\alpha n}(x)|^2}{\hbar |v_{\alpha n}|},
\end{equation}
where $f_0$ is the zero-bias distribution function, $\Psi_{\alpha n}(x)$ and \( v_{\alpha n} \) are
the wave function and velocity corresponding to the incident mode \( n \) from terminal \( \alpha \).
The total injectivity contains the contributions from all terms as
\begin{equation}
    \frac{dn}{dE}(x) = \sum_{\alpha}\frac{dn_{\alpha}}{dE} (x).
\end{equation}
In the weakly nonlinear regime, we expand the current to the second-order of the bias voltages as~\cite{Christen1996,Wang1999,Wei2022}
\begin{equation}\label{2ndIeq}
I_\alpha=\sum_\beta G_{\alpha \beta}\mathcal{V}_{\beta}+
\sum_{\beta \gamma} G_{\alpha \beta \gamma} \mathcal{V}_\beta \mathcal{V}_\gamma+\cdots.
\end{equation}

\section{nonreciprocal transport in the ballistic regime}
In this section, we employ the theoretical framework
introduced in the previous section to study nonreciprocal ballistic transport.
We here focus on transport in a two-terminal setup. The current is assumed to flow
in the \(x\) direction. The two terminals are situated on the left (L) and right (R) sides, with
the corresponding biases denoted by \( \mathcal{V}_{\text{L}} \) and \( \mathcal{V}_{\text{R}} \), respectively.
Due to the gauge invariance, the current flowing in two-terminal setups
as formulated in Eq.~\eqref{2ndIeq}, solely depends on the bias difference between the two terminals,
which can be described by
\begin{equation}\label{I2terminal}
    I(\Delta V) = G_1 \Delta V + G_2 \Delta V^2 + \mathcal{O}(\Delta V^3),
\end{equation}
where \( \Delta V = \mathcal{V}_\text{L} - \mathcal{V}_\text{R} \).
For convenience, we set the bias configuration as \( \mathcal{V}_{\text{L}} = \mathcal{V}/2 \)
and \( \mathcal{V}_{\text{R}} = -\mathcal{V}/2 \),
corresponding to a voltage difference \( \Delta V = \mathcal{V} \).
Accordingly, the current in Eq.~\eqref{Currenteq} reduces to
\begin{equation}\label{I2tLBEq}
    \begin{split}
        I (\mathcal{V}) = &-\frac{e}{h} \int dE \,
        T(E, U(\mathcal{V}))\times \\
        &\left[f\left(E-E_F+\frac{e\mathcal{V}}{2}\right) - f\left(E-E_F-\frac{e\mathcal{V}}{2}\right)\right].
    \end{split}
\end{equation}
Without the bias induced Coulomb potential, it is straightforward to
prove the reciprocity of the transport, that is \( I(\mathcal{V}) = -I(-\mathcal{V}) \).
This conclusion fails as the bias dependent Coulomb potential \( U(\mathcal{V}) \)
is taken into account. Eq.~\eqref{Uexpandeq} now reduces to
\begin{equation}\label{U1stEq}
    U(\mathcal{V}) = \left(u_{\text{L}} - u_{\text{R}}\right)\frac{\mathcal{V}}{2} + \mathcal{O}(\mathcal{V}^2),
\end{equation}
to the first order of \( \mathcal{V} \). From Eq.~\eqref{I2tLBEq}, it is evident that if \(T(E, U(\mathcal{V})) \neq T(E, U(-\mathcal{V}))\)
then \(|I(\mathcal{V})| \neq |I(-\mathcal{V})|\), giving rise to nonreciprocal ballistic transport.
This can be achieved by \( U(\mathcal{V}) \neq U(-\mathcal{V}) \), or \( u_{\text{L}} \neq u_{\text{R}} \)
in Eq.~\eqref{U1stEq}. According to Eq.~\eqref{cpueq}, this means that \( dn_{\text{L}}/dE \neq dn_{\text{R}}/dE \).
One way to achieve this condition is by breaking the translational symmetry along the transport direction using proper device geometries ~\cite{Christen1996,Sheng1998,Song1998}. However, this strategy
can readily drive the system away from the ballistic regime.

Here, we propose an alternative approach to realize nonreciprocal ballistic transport
taking advantage of the asymmetrical band structures with $E(k_x^{\text{L}})\neq E(k_x^{\text{R}})$,
where $k_x^{\text{L,R}}$ are the wave vector for the left and right-moving state, respectively.
The scenario is to lift the condition of
\( U(\mathcal{V}) = U(-\mathcal{V})\)
by introducing unequal LPDOS in Eq.~\eqref{LPDOSPSiEq}
for different terminals, which is achieved by the bias resolved Coulomb potential.
Specifically, for a given energy $E$, the opposite propagation states possess unequal velocities,
\begin{equation}
    \left|v(k_{x}^{\text{L}}, E)\right| \neq \left|v(k_{x}^{\text{R}}, E)\right|,
\end{equation}
due to the asymmetry of the bands.
For a uniform system, the spatial distribution of the eigenstates \( |\Psi_{\alpha n}(x)| \)
is independent of \( x \), and so is the LPDOS \( dn_{\alpha}/dE \).
Similarly, the Coulomb potential remains constant throughout the scattering region
so that the characteristic potentials satisfy \( \nabla_x^2 u_{\alpha} = 0 \).
The characteristic potentials are then expressed as
\begin{equation}\label{cpuSoleq}
    u_{\alpha} = \frac{dn_{\alpha}}{dE}\bigg/\frac{dn}{dE}.
\end{equation}
This result is consistent with the local neutral approximation~\cite{Christen1996,Kramer2012,Wang1999},
where the local charge density is assumed to be zero everywhere inside the system.
Since the LPDOS is solely determined by the velocities in Eq.~\eqref{LPDOSPSiEq},
the asymmetric band structures assign different values to the LPDOS for the two terminals
with \( dn_{\text{L}}/dE \neq dn_{\text{R}}/dE \),
which further gives $u_{\text{L}}\neq u_{\text{R}}$.

The analysis above highlights the key scenario for the
nonreciprocal ballistic transport in asymmetric bands.
Next, we introduce \( U_{c} \equiv U(\mathcal{V}) = -U(-\mathcal{V}) \), and
drop the higher order terms \( \mathcal{O}(\mathcal{V}^2) \) in Eq.~\eqref{U1stEq},
and denote the transmission for \( +\mathcal{V} \) as \( T_{+} \equiv T_{LR}(E, U_c) \)
and that for \( -\mathcal{V} \) as \( T_{-} \equiv T_{LR}(E, -U_c) \) for brevity.
It is conceivable that the nonreciprocal condition \( T_{+} \neq T_{-} \) holds in general.
Expanding Eq.~\eqref{I2tLBEq} in the weak nonlinear regime yields
\begin{equation}\label{2ndConsUcIeq}
    \begin{split}
        I\left(\mathcal{V}\right) &\approx \frac{e^2}{h}\int dE \left(-\partial_E f_0\right)
        \left[T_0 \mathcal{V} + \frac{e}{2} \left(\partial_E T_0\right)
        \left(u_{\text{L}} - u_{\text{R}}\right) \mathcal{V}^2 \right]\\
        &= G_1 \mathcal{V} + G_2 \mathcal{V}^2,
    \end{split}
\end{equation}
where \( T_0 \equiv T_{\text{LR}}(E, U=0) \),
and the first-order and second-order conductances are expressed as
\begin{equation}\label{G1G2bal}
    \begin{split}
        G_1 &= \frac{e^2}{h} \int dE  \left(-\partial_E f_0\right) T_0,\\
        G_2 &= \frac{e^3}{2h} \left(u_{\text{L}} - u_{\text{R}}\right)\int dE \left(-\partial_E f_0\right)
        \left(\partial_E T_0\right).
    \end{split}
\end{equation}
In the limit of zero temperature, the integration in Eq.~\eqref{2ndConsUcIeq} simplifies to
\begin{equation}\label{2ndConsUcI0Keq}
    I \approx \frac{e^2}{h}
    \left[T_0 \mathcal{V} +
        \frac{e}{2} \left(\partial_E T_0\right)
    \left(u_{\text{L}} - u_{\text{R}}\right) \mathcal{V}^2 \right],
\end{equation}
where the energy is assumed to be at the Fermi energy \( E_F \).

\section{Nonreciprocal ballistic transport in Rashba electron gas}

The scenario of nonreciprocal ballistic transport introduced in the previous section is general.
In this section, we elucidate the physical effects using the concrete example of 2D Rashba electron gas
subjected to a in-plane magnetic field, which realize asymmetric band structures.
We investigate the modulation of the internal Coulomb potential induced by the bias voltages
and the resultant nonreciprocal transport properties. The physical conditions for
the implementation of the nonreciprocal transport are given.

\subsection{2D Rashba gas with asymmetric bands}

\begin{figure}[t]
	\centering
	\includegraphics[width=1\columnwidth]{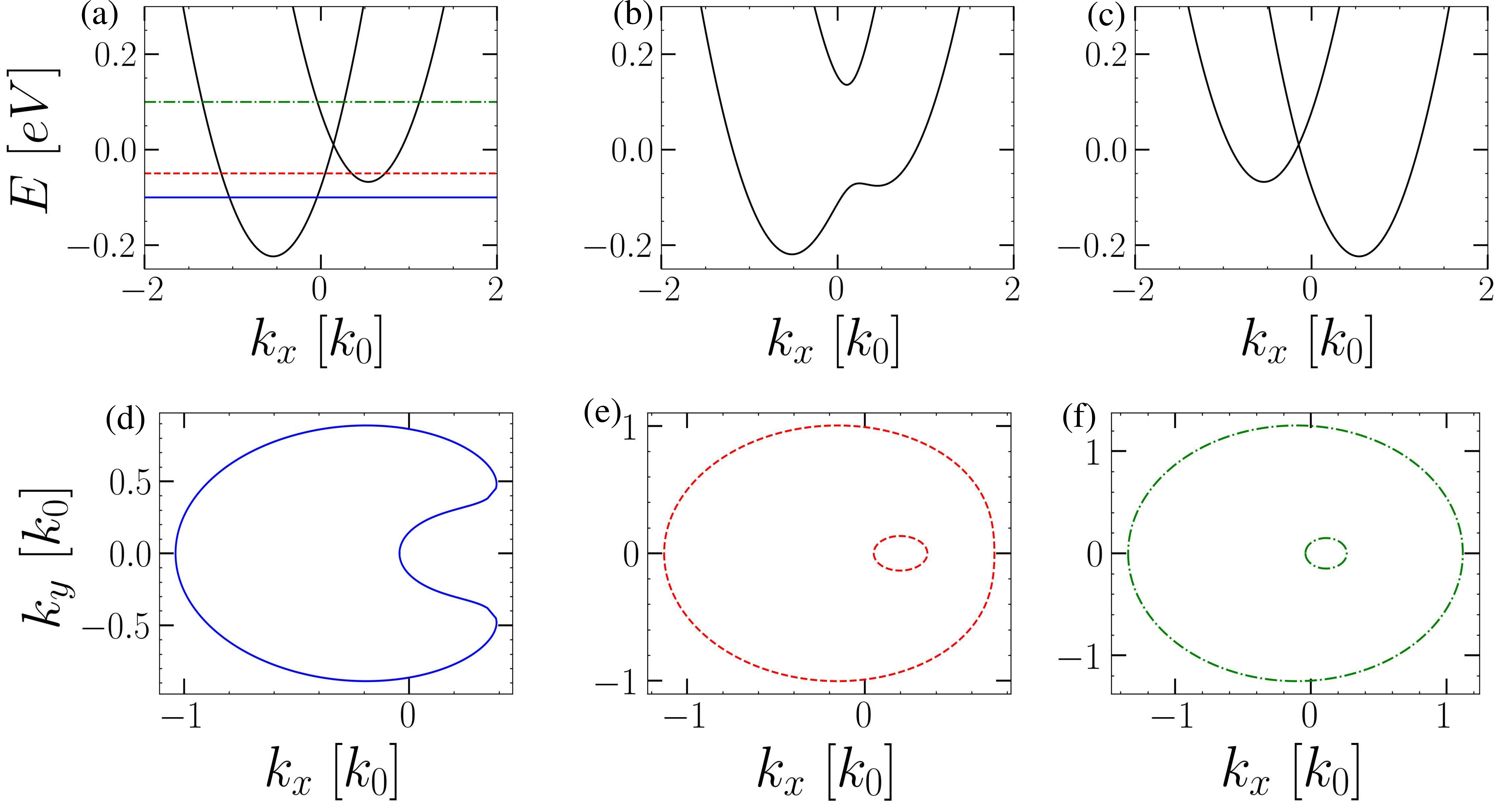}
	\caption{The energy spectrum of the 2D SOC system, as calculated from Eq.~\eqref{Ebandeq}, is illustrated in the following configurations: (a) and (c) display the energy spectrum of the SOC system with \( B_y = B_0 \) and \(- B_0 \), respectively, at a fixed \( k_y = 0 \). (b) shows the energy spectrum with \( B_y = B_0 \) and \( k_y = 0.2 k_0 \). (d-f) depict the Fermi surfaces, which are closed contours in the 2D system, at \( E_F = -0.1, -0.05, 0.1 \, \text{eV} \), corresponding to the levels shown in (a), with \( B_y = B_0 \).
		Here, the parameters are \( m = 0.15 m_e \), with \( m_e \) being the mass of a bare electron, \( \lambda = 3.85 \, \text{eV} \, \text{\r{A}} \), \( k_0 \) satisfying \( \hbar^2 k_0^2/m = 1 \, \text{eV} \), and \( B_0 = 0.078 \, \text{eV} \).
	}\label{fig1}
\end{figure}

We consider the 2D electron gas with Rashba spin-orbit coupling (SOC) subjected to
an in-plane magnetic field, which can be captured by the Hamiltonian as
\begin{equation}\label{2DSOCEq}
    H\left(\mathbf{k}\right) = \frac{\hbar^2 k^2}{2m} + \lambda\left(k_x \sigma_y - k_y \sigma_x\right) - B_y \sigma_y,
\end{equation}
where \( m \) denotes the effective electron mass, \( \lambda \) represents the strength of the SOC,
\( k^2 = k_x^2 + k_y^2 \) defines the magnitude of the wave vector, and \( B_y \) is the
Zeeman splitting induced by the in-plane magnetic field along the \( y \)-axis.

When \( B_y \) is zero, the system satisfies both space inversion symmetry (SIS) and
time reversal symmetry (TRS). Utilizing the Qsymm package\cite{Varjas2018}, we
identify the unitary symmetries related to SIS, including inversion symmetry with the action \( \mathcal{S}=\sigma_z \), and mirror symmetries along \( x \) and \( y \) directions with actions \( \mathcal{M}_x=\sigma_x \) and \( \mathcal{M}_y= i \sigma_y \), respectively. These symmetry operations transform the Hamiltonian as follows:
\begin{equation}
    \begin{split}
        \mathcal{S}^{\dagger} H(\mathbf{k}) \mathcal{S} &= H(-\mathbf{k}), \\
        \mathcal{M}^{\dagger}_x H(k_x, k_y) \mathcal{M}_x &= H(-k_x, k_y), \\
        \mathcal{M}^{\dagger}_y H(k_x, k_y) \mathcal{M}_y &= H(k_x, -k_y).
    \end{split}
\end{equation}
TRS is represented as an antiunitary symmetry with the action \( \mathcal{T}=\sigma_y \mathcal{K} \), where \( \mathcal{K} \) is the complex conjugation operator, and it satisfies:
\begin{equation}
    \mathcal{T}^{\dagger} H(\mathbf{k}) \mathcal{T} = H(-\mathbf{k}).
\end{equation}
When \( B_y \) is nonzero, symmetries associated with \( \mathcal{S} \), \( \mathcal{M}_x \), and \( \mathcal{T} \) are all
broken, leading to an asymmetric band structure along the \( x \) direction
(cf. Fig.~\ref{fig1}). Specifically, the energy dispersions are
\begin{equation}\label{Ebandeq}
    E_{\pm}(\mathbf{k}) = \frac{\hbar^2 k^2}{2m} \pm
    \sqrt{ (\lambda k_x - B_y)^2 + \lambda^2 k_y^2 }.
\end{equation}
The $x$-component of the velocities are
\begin{equation}
    v^x_{\pm} = \frac{\hbar k_x}{m} \pm \frac{\lambda\left( \lambda k_x - B_y \right)}
    {\hbar \sqrt{\left(\lambda k_x - B_y \right)^2 + \lambda^2 k_y^2}}.
\end{equation}

\begin{figure*}[t]
	\centering
	\includegraphics[width=1.8\columnwidth]{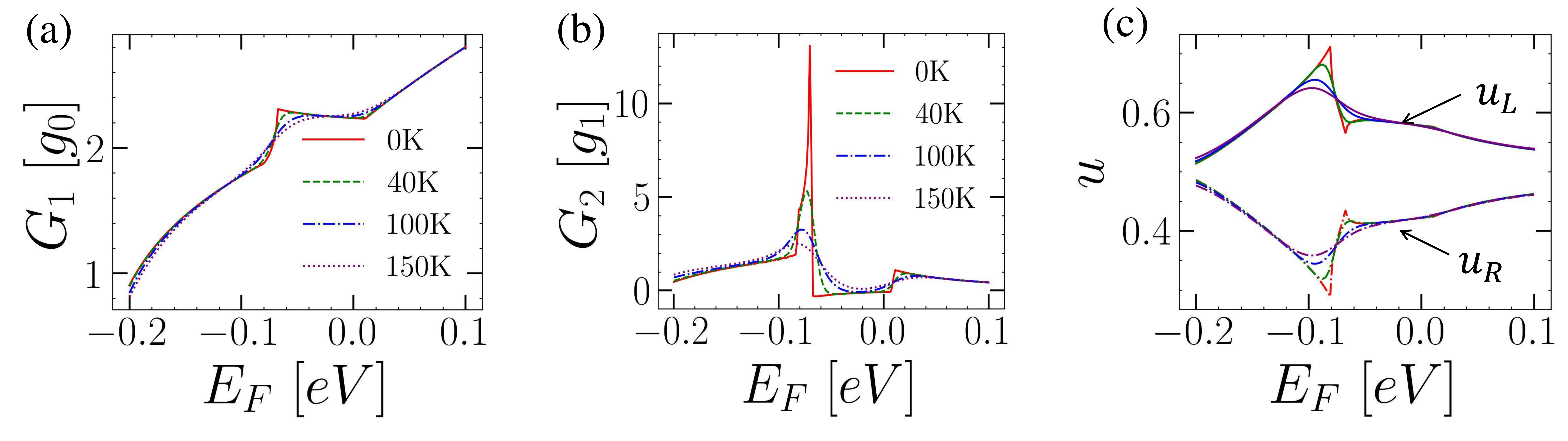}
	\caption{Figures (a), (b), and (c) display the first-order conductances, the second-order conductances, and the characteristic potentials of the two terminals, respectively, at varying temperatures of $0\,$K, $40\,$K, $100\,$K, and $150\,$K. In these figures, \( B_y = B_0 \) is fixed, and the other system parameters are identical to those presented in Fig.~\ref{fig1}. Specifically, in Fig. (c), the solid and dashed lines represent the characteristic potentials \( u_L \) and \( u_R \), respectively. These lines correspond to the same temperatures as in Figs. (a) and (b), indicated by matching colors. The parameters \( g_0 \) and \( g_1 \) are defined such that \( g_0 = \frac{e^2 L_y k_0}{2\pi h} \) and \( \frac{g_0}{2g_1} = 1\,\)V.
	}\label{fig2}
\end{figure*}

Here, the potential \( U \) is assumed to be independent of \( y \), predicated
on the uniformity of the system along the \( y \)-direction. This assumption facilitates a simplified solution for the characteristic potential, analogous to the formulation presented in Eq.~\eqref{cpuSoleq}.
The LPDOS contains the contribution from all $k_y$ channels and can be expressed as
\begin{equation}
    \frac{dn_{\alpha}}{dE} = \int \frac{dk_y}{2\pi} \frac{dn_{\alpha, k_y}}{dE},
\end{equation}
where \( dn_{\alpha, k_y}/dE \) represents the LPDOS of the \( k_y \) channel in
the terminal \( \alpha \). In the ballistic limit, where \( k_x \) remains a good quantum
number and the terminals share the same Hamiltonian as the system, the total transmission is directly evaluated as
\begin{equation}\label{Tbal}
    T_0 = \frac{L_y}{2\pi} \sum_{n=\pm}\int d k_x d k_y \theta\left(v^x_n\right) v^x_{n} \delta\left(E-E_n\right),
\end{equation}
where \( L_y \) denotes the width of the scattering region in the \( y \) direction,
\( \theta\left(\cdots\right) \) and \( \delta\left(\cdots\right) \) are
the Heaviside step function and Dirac delta function, respectively.
The LPDOS of the left and right terminals can be simplified as
\begin{equation}\label{lpdosbal}
    \frac{dn_{L,R}}{dE} = \frac{-1}{\left(2\pi\right)^2} \sum_{n=\pm}\int d k_x d k_y \theta\left(\pm v^x_n\right) \partial_{E_n} f\left(E_n-E_F\right),
\end{equation}
where \( L, R \) correspond to \( +v^x_n, -v^x_n \) respectively.
As the temperature approaches zero, \( dn_{L,R}/dE \) becomes proportional to the
carrier density of the Fermi arcs for the left- and right-moving modes, as indicated
by Eq.~\eqref{lpdosbal}. When the Fermi surfaces (loops in the 2D case) are asymmetric,
as shown in Figs.~\ref{fig1}(d-f), \( dn_{L}/dE \neq dn_{R}/dE \). According to Eqs.~\eqref{cpuSoleq}
and \eqref{2ndConsUcIeq}, nonreciprocal transport phenomena can be expected.

\begin{figure}[t]
	\centering
	\includegraphics[width=1\columnwidth]{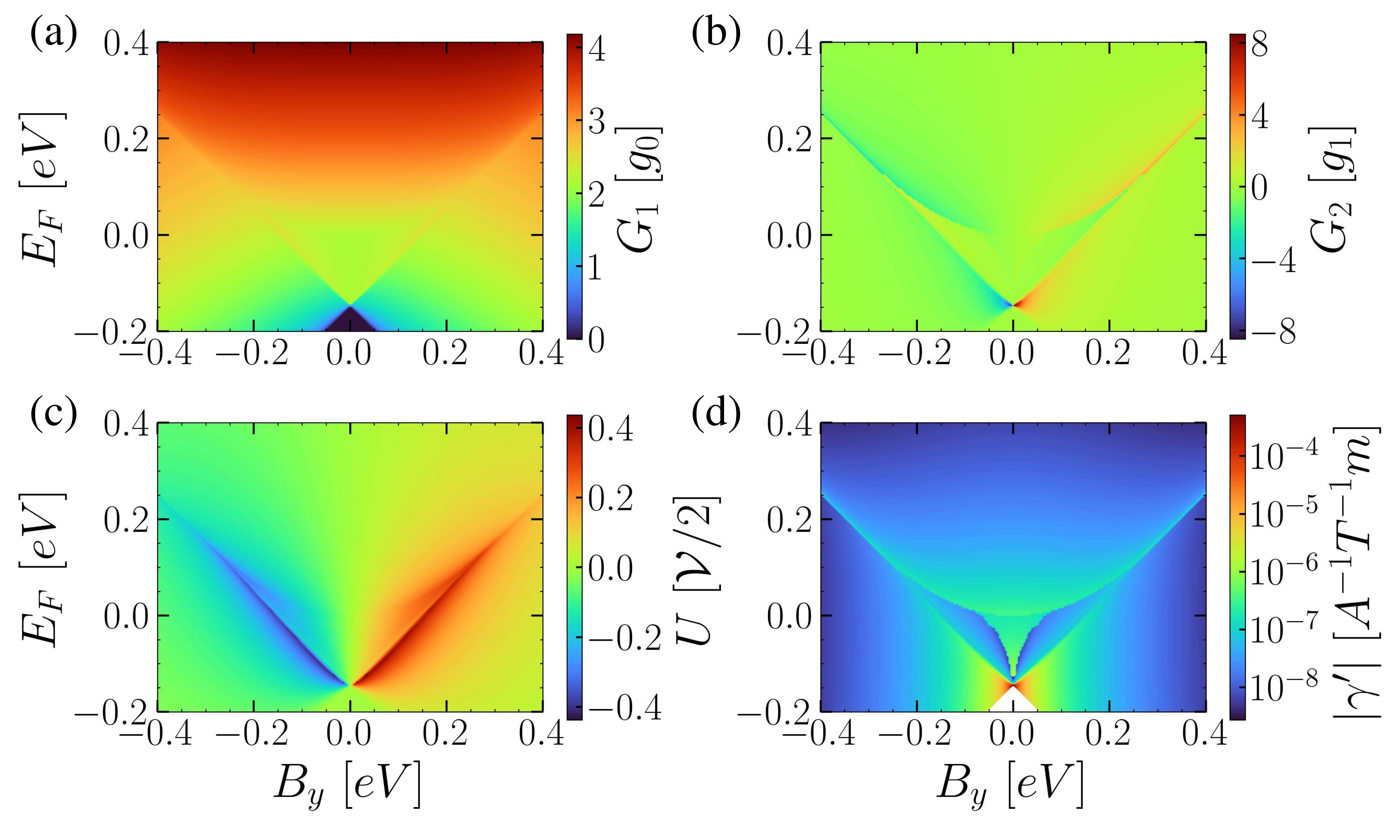}
	\caption{Contour plot figures are presented to illustrate:
		(a), (b) the first-order conductance \( G_1 \) and second-order conductance
		\( G_2 \), as expressed in Eq.~\eqref{G1G2bal},
		(c) the first-order potential \( U \), as expressed in Eq.~\eqref{U1stEq}, and
		(d) the magnitude of  $\gamma^{\prime}=\gamma L_y$ with $\gamma$ defined in Eq.~\eqref{ResistenceEq}.
		They are in relation to variations in the Fermi energy and the Zeeman energy \( B_y \).
		The parameters here are consistent with those used in Fig.~\ref{fig2}, with the temperature fixed at \( 0 \, \)K.
		We take $g=60$ as the $g$-factor to determine the magnetic field when calculating $\gamma$.
	}\label{fig3}
\end{figure}

\subsection{Results and Discussions}

We start by analyzing the asymmetric energy bands described by Eq.~\eqref{Ebandeq}, as depicted in Fig.~\ref{fig1}.
For a given energy in the $k_y$ channel, the velocities of the counter-propagating states are unequal,
so as the density of states (DOS).
Such band asymmetry necessitates the violation of SIS and TRS which are symmetries $\mathcal{S}$, $\mathcal{M}_x$ and $\mathcal{T}$ here.
Figs.~\ref{fig1}(d-f) reveal the broken reflection symmetry of the Fermi loops about the $k_x$ axis,
which gives rise to different characteristic potentials $u_L$ and $u_R$
and leads to the emergence of second-order conductance $G_2$, as formulated in Eq.~\eqref{G1G2bal}.
In Fig.~\ref{fig2}, we plot the characteristic potentials and two conductances $G_{1,2}$.
The nonreciprocal transport is revealed by the second-order conductance $G_2$ in Fig.~\ref{fig2}(b).
According to Eq.~\eqref{G1G2bal}, it is proportional to the difference of characteristic potentials $u_{\text{L}}-u_{\text{R}}$;
see Fig.~\ref{fig2}(c). From Fig.~\ref{fig2}(b), one can see that
the most pronounced nonreciprocal signals occur at $E_F\approx-0.07\,$eV,
when the Fermi energy coincides with the higher band bottom in Fig.~\ref{fig1}(a).
It is associated with the von Hove singularity in the density of states (DOS), resulting in a sudden alteration in conductance $G_1$, as depicted in Fig. \ref{fig2}(a), where the derivative of $G_1$ undergoes an abrupt change at $E_F\approx-0.07\,$eV. This observation aligns with findings in the diffusive limit \cite{Ideue2017}.
However, our results in the ballistic limit display persistent non-vanishing signals even when the Fermi energy is above the Dirac point, thus differentiating them from diffusive transport behaviors \cite{Ideue2017}.

In Fig.~\ref{fig3}, we illustrate how the conductances and first-order potential are influenced by both the Fermi energy and the magnetic field $B_y$.
When $B_y$ is inverted to $-B_y$, the conductance $G_1$ remains unchanged, as does its derivative with respect to energy; see Fig.~\ref{fig3}(a).
Concurrently, the energy band structure reverses along the $k_x$ direction, as shown in Figs.~\ref{fig1}(a) and (c).
This reversal swaps the characteristic potentials from $u_{L(R)}$ to $u_{R(L)}$, given that the LPDOS
is fully determined by the band structures.
This phenomenon is manifested in Fig.~\ref{fig3}(c), where the first-order potential $U$ is an odd function of $B_y$ for a given $E_F$.
As for Eq.~\eqref{G1G2bal}, the conductance $G_2$ should also invert its sign, corresponding to the observation in Fig.~\ref{fig3}(b).
It is noted that the Onsager reciprocal relation~\cite{Onsager1931} is violated, as $|I(\mathcal{V}, B)| \neq |I(\mathcal{-V}, B)|$.
However, the relation $|I(\mathcal{V}, B)| = |I(-\mathcal{V}, -B)|$ holds, akin to the EMCA effects~\cite{Rikken2001,Krstic2002,Rikken2005}.

The resistance formula is commonly adopted in nonreciprocal transport measurements.
The expression for resistance, as discussed in Refs.~\onlinecite{Rikken2001,Rikken2005}, can be obtained from Eq.~\eqref{2ndConsUcIeq} as
\begin{equation}\label{ResistenceEq}
    R = \frac{\mathcal{V}}{I} = R_0\left(1 - \gamma I B \right),
\end{equation}
where
\begin{equation*}
    R_0 = \frac{1}{G_1}, \quad
    \gamma = \frac{G_2}{G_1^2 B}.
\end{equation*}
We plot $\gamma^{\prime}=\gamma L_y$ as a function of $B_y$ and $E_F$ in Fig.~\ref{fig3}(d).
In the funnel-mouth-shaped region, where $B_y$ is relatively small and the Fermi energy lies between the higher band bottom and the Dirac point, $\gamma^{\prime}$ appears to be largely independent of $B_y$.
This result coincides with that in the diffusive limit~\cite{Ideue2017}.
However, it is noteworthy that the magnitude of $\gamma^{\prime}$ in this region, approximately $10^{-6}\,$A$^{-1}$T$^{-1}$m, is significantly larger than that reported in Ref. \onlinecite{Ideue2017}.

\section{conclusions}
In conclusion, our theoretical investigation focuses on nonreciprocal transport in energy band asymmetric systems within the quantum ballistic regime.
A pivotal aspect of our study is the consideration of Coulomb potentials induced by finite biases.
Antisymmetric biases at the left and right terminals lead to asymmetric potentials, a consequence of the inherent asymmetry in the band structure.
Additionally, our analysis anticipates significantly larger nonreciprocal current signals in the quantum transport regime compared to diffusive bulk materials.
We hope that our work will provide valuable insights for future quantum transport experiments.

\begin{acknowledgments}
M. H. Zou and H. Geng contributed equally to this work.

This work was supported by  the State Key Program for Basic Researches of
China under Grants No. 2021YFA1400403 (D.Y.X.),
and the National Natural Science Foundation of China under Grant
No. 11974168 (L.S.),
No. 12174182 (D.Y.X.),
No. 12074172 (W.C.),
No. 12222406 (W.C.),
No. 12274235 (R.M.),
and No. 12304068 (H. G.).
\end{acknowledgments}

\bibliographystyle{apsrev4-1}

\bibliography{nonlinear}

\end{document}